\documentclass[pre,twocolumn,reprint]{revtex4-1}
\usepackage{amsmath,amsfonts,amssymb}
\usepackage{ucs}
\usepackage[utf8x]{inputenc}
\usepackage[english]{babel}
\usepackage{fontenc}
\usepackage{graphicx}
\usepackage{subfigure}
\usepackage[dvips]{hyperref}

\begin{document}

\title{Stream of asymmetric bubbles in a Hele-Shaw channel}

\author{Ant\^onio M\'arcio~P.~Silva}
\affiliation{Laborat\'orio de F\'\i sica Te\'orica e Computacional,
Departamento de F\'{\i}sica, Universidade Federal de Pernambuco,
50670-901 Recife, PE, Brazil}
\author{Giovani L.~Vasconcelos}
\email{giovani@df.ufpe.br}
\affiliation{Laborat\'orio de F\'\i sica Te\'orica e Computacional,
Departamento de F\'{\i}sica, Universidade Federal de Pernambuco,
50670-901 Recife, PE, Brazil}

\begin{abstract}
Exact solutions are reported for a stream of  asymmetric bubbles steadily moving in a Hele-Shaw channel. From the periodicity along the streamwise direction, the flow region is reduced to a rectangular unit cell containing one bubble, which is conformally mapped to an annulus in an auxiliary complex plane. Analytic expressions for the bubble shape as well as for the velocity field are obtained in terms of the generalized Schwarz-Christoffel formula for doubly connected domains. 
\end{abstract}

 \maketitle

\section{Introduction}

The study of interface dynamics in a Hele-Shaw
cell, where the fluids are confined between two closely spaced glass
plates, has been a topic of great interest since the seminal work by Taylor and Saffman \cite{ST}.  
More recently, the Hele-Shaw system  was
shown to have an integrable structure and to possess deep
connections with other branches of mathematical physics \cite{MWZ}.
This rich mathematical structure lies behind the fact that 
numerous exact solutions have been obtained both for time-dependent
interfaces (listed in \cite{Gustafsson}) and for steady shapes 
(see \cite{GLV2001} and references therein). 
Most of these exact solutions  were obtained for situations where the flow domain is (or can be mapped to) a simply connected region. Hele-Shaw flows in multiply connected regions remain a problem of considerable difficulty and only a few solutions are known \cite{richardson1994,crowdy2009a,pof2012}. One particular instance where multiply connected fluid domains naturally appear is in the case of multiple steady bubbles in a Hele-Shaw channel.
 This class of problems, besides being mathematically challenging, is also of practical interest in connection, for instance, with the problem of clustering in bubbly Hele-Shaw  flows  \cite{zenit2001,zenit2005}.  Periodic arrays of bubbles in a Hele-Shaw channel are also of interest in view of similar arrangements found in the flow of vesicles 
and red blood cells  in microcapillaries \cite{blood1,blood3,blood2}. 
For example, the ``zigzag'' phase \cite{blood1,blood2} observed in the flow of  red blood cells in microvessels has a counterpart in the Hele-Shaw system as a staggered two-file array of bubbles \cite{prsa2011}. 

In the present paper we report exact solutions for a periodic stream of  asymmetric bubbles steadily moving in a Hele-Shaw channel. Because of the lack of  symmetry, one has to deal with a fluid domain (within a unit cell) that  is doubly connected. We tackle the problem by conformally mapping the fluid domain to an annulus in an auxiliary complex plane. An explicit solution for the bubble shape and the corresponding velocity potential is then obtained in terms of the generalized Schwarz-Christoffel formula for doubly connected domains, and several specific examples of bubble configurations are discussed. We emphasize that this is the first instance of exact solutions for completely asymmetric bubbles in a Hele-Shaw channel.  

It is worth noting that exact solutions for a single asymmetric bubble in a Hele-Shaw channel have been obtained by using either conformal mapping techniques \cite{tanveer87} or  Riemann-Hilbert methods \cite{CD1988}.  In this case, however, the bubble has fore-and-aft symmetry and so the flow region can effectively be reduced to a simply connected domain \cite{glv2001}.  Solutions for multiple steady bubbles in an unbounded Hele-Shaw cell were reported in Ref.~\cite{crowdy2009a}; here the bubbles also had fore-and-aft symmetry but the flow domain was multiply connected. An attempt at extending these solutions to include bubbles with no assumed symmetry in a channel was made in Ref.~\cite{crowdy2009b} but it was subsequently found that part of the argument used there was not valid \cite{crowdy}.  On the other hand, it seems possible to extend the method presented herein to the case of a finite number of asymmetric bubbles in a Hele-Shaw channel. This requires conformal mappings between 
domains of arbitrary connectivity 
and will be presented elsewhere \cite{GV2013}.

\section{Mathematical Formulation}

\begin{figure}[t]
\centering 
\subfigure[\label{fig:2a}]{\includegraphics[width=5cm]{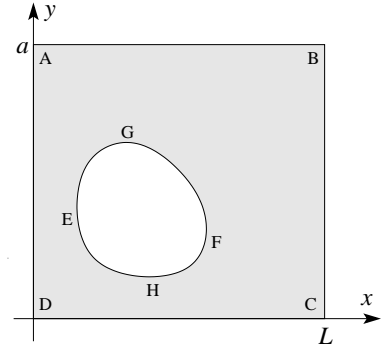}}
\subfigure[\label{fig:2b}]{\includegraphics[width=5cm]{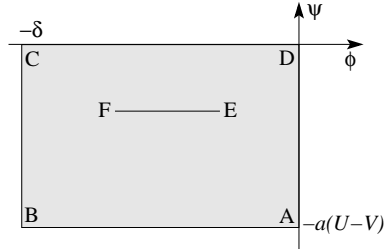}}
\subfigure[\label{fig:2c}]{\includegraphics[width=5cm]{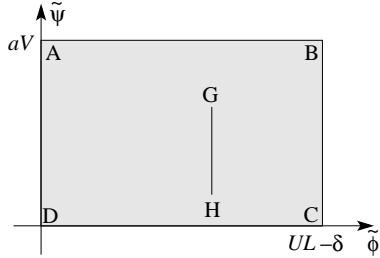}}
\subfigure[\label{fig:2d}]{\includegraphics[width=5cm]{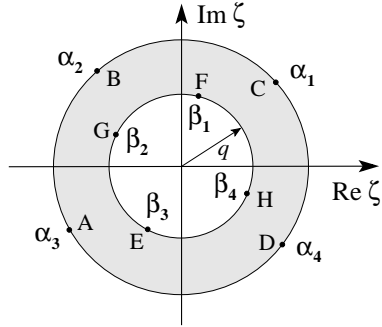}}
\caption{The flow domains for a reduced unit cell: (a) the $z$-plane, (b) the w-plane, (c) the $\tilde w$-plane, and (c) the auxiliary $\zeta$-plane.}\label{fig:2}
\end{figure}

We consider the problem of a periodic stream of bubbles moving
with a constant velocity $U$ along the $x$-direction in a
Hele-Shaw channel of width $a$. As usual, we will work in a frame co-moving with the bubbles.  The array of bubbles has a spanwise  period denoted by $2L$, and in each period cell there are two bubbles that are mirror reflections of one another with respect to a midline perpendicular to the channel walls, chosen as the $y$ axis.  This ensures that both  the $y$ axis and the lateral edges, $y=\pm L$, of the period cell are equipotentials of the flow, which implies in turn that the problem can be 
reduced to a rectangular unit cell corresponding to one half, say the right half, of the original period cell; see
Fig.~\ref{fig:2a}.

We recall that the motion of a viscous fluid in a Hele-Shaw cell is governed by Darcy's law, $\vec{v}=-({b^2 }/{12\mu})\vec{\nabla} p$, where $b$ is the gap between the
cell plates, $\mu$ is the viscosity, and $p$ is the fluid pressure. Since the fluid is incompressible, i.e., $\vec{\nabla}\cdot\vec{v}=0$, it follows that $p$ is a harmonic function $\nabla^2 p=0$. It is thus convenient to introduce the complex potential   $w(z) = \phi +\rm i \psi$, where $z=x+\rm i y$, $\phi=-({b^2 }/{12\mu})p -Ux$ is the velocity potential in the co-moving frame, and $\psi$ is the associated  stream function. In terms of the complex potential $w(z)$ the problem can be formulated  as follows. Now let $\cal D$ be the domain occupied by the viscous fluid within the reduced cell and let $\cal C$ be the 
bubble interface. The complex potential $w(z)$ must then be analytic in the fluid domain $\cal D$ and satisfy the following boundary conditions on $\mathcal C$: 
\begin{align}
w=-Ux+\mbox{constant},
 \label{eq:cc1}
\end{align}
which follows from the combined facts that the bubble surface $\mathcal C$ is a streamline of the flow, the pressure inside the bubble is constant (taken to be zero), and surface tension effects are neglected.

Additional boundary conditions follow from the fact  that the upper and lower edges of the unit cell are streamlines of the flow, that is,
\begin{align}
\text{Im}~w=-a(U-V) \qquad \text{on} \qquad y = a, \label{eq:cc02}
\end{align}
\begin{align}
\text{Im}~w=0 \qquad \text{on} \qquad  y=0, \label{cc.03a}
\end{align}
whereas the left and right edges of the unit cell are equipotentials:\begin{align}
\text{Re}~w=0 \qquad \text{on} \qquad x=0, \label{cc.04a}
\end{align}
\begin{align}
\text{Re}~w=-\delta \qquad \text{on} \qquad x= L, \label{eq:cc05}
\end{align}
where $\delta$ is a real-valued constant whose value is related to the period $L$ and to the bubble area \cite{prsa2011}. The flow domain in the $w$-plane thus corresponds to a rectangle with a horizontal slit in its interior, as shown in Fig. \ref{fig:2b}.

To solve the free-boundary problem formulated in Eqs.~(\ref{eq:cc1})--(\ref{eq:cc05}), it is convenient to introduce the  following function:
\begin{align}
\tilde{w}(z)=w(z)+Uz.
\label{eq:tw}
\end{align}
which corresponds to the complex potential in the lab frame.
From the definition (\ref{eq:tw}) and Eqs.~(\ref{eq:cc1})--(\ref{eq:cc05}), it follows that $\tilde{w}(z)$  satisfies the following boundary conditions:
\begin{align}
 \tilde w = iUy+C \qquad \text{on} \qquad \cal C, \label{cc.01b}
\end{align}
\begin{align}
 \text{Im}~\tilde w= aV \qquad \text{on} \qquad y= a, \label{cc.02b}
\end{align}
\begin{align}
 \text{Im}~\tilde w=0 \qquad \text{on} \qquad y=0, \label{cc.03b}
\end{align}
\begin{align}
 \text{Re}~\tilde w=0 \qquad \text{on} \qquad  x = 0, \label{cc.04b}
\end{align}
\begin{align}
 \text{Re}~\tilde{w}=UL-\delta \qquad \text{on} \qquad  x= L. \label{cc.05b}
\end{align}
From these conditions, one sees that the flow domain in the $\tilde w$-plane is a rectangle with a vertical slit in its interior, as shown in Fig. \ref{fig:2c}.

Let us now consider the conformal mapping $z(\zeta)$ from an annulus, $q\le|\zeta|\le1$,  in an auxiliary complex $\zeta$-plane onto the fluid domain in the $z$-plane, in such a way that the  inner circle, $|\zeta|=q$, is mapped to the bubble $\cal C$ and the unit circle, $|\zeta|=1$, is mapped to the edges of  our unit cell in the $z$-plane;  see Fig. \ref{fig:2d}. The corners of the rectangular cell in the $z$-plane
are  the pre-images of four points, denoted by $\alpha_i$, on the unit circle (i.e., $|\alpha_i|=1$, for $i=1,2,3,4$). Similarly,   the rightmost, leftmost,  topmost, and lowermost points on the bubble are the pre-images of four  points  $\beta_i$  on the inner circle (i.e., $|\beta_i|=q$,  for $i=1,2,3,4$), as indicated in Fig.~\ref{fig:2}.

Defining the functions
\begin{align}
W(\zeta)=w(z(\zeta)) \qquad \mbox{and} \qquad \tilde W(\zeta)=\tilde w(z(\zeta)), 
\end{align}
it follows from (\ref{eq:tw}) that  the conformal mapping 
$z(\zeta)$ can be written as
\begin{equation}
 z(\zeta)=\frac{1}{U}\left[\tilde W(\zeta)- W(\zeta)\right]. \label{eq:z}
\end{equation}
We have thus reduced our original free-boundary problem to the more amenable problem of obtaining two analytic functions,  $W(\zeta)$ and $\tilde W(\zeta)$,  that map an annulus onto respective rectangular domains with linear slits in their interior. The latter problem can conveniently be handled in terms of   the generalized Schwarz-Christoffel formula for doubly connected domains, as discussed below. Note furthermore that once the conformal mapping $z(\zeta)$ is known,  the bubble shape can be readily computed  in parametric form by setting: $x(\theta)+iy(\theta)=z(qe^{i\theta})$, for $0<\theta<2\pi$.

\section{The General Solution}

\subsection{The function $W(\zeta)$}

Since the conformal mapping $w=W(\zeta)$  maps an  annulus in the $\zeta$-plane onto a (degenerate) polygonal domain in the $w$-plane, an explicit solution for this mapping can be obtained by a direct application of  the Schwarz-Christoffel formula for doubly connected regions \cite{crowdy1}. In this case, one finds that $W(\zeta)$ is given by the following expression:
\begin{equation}
W(\zeta)={\mathcal A} + {\mathcal K}\int_{\zeta_0}^\zeta \frac{P(q^2\zeta/\beta_1,q)P(q^2\zeta/\beta_3,q)}
{\sqrt{\prod_{i=1}^4P(\zeta/\alpha_i,q)}}\, d\zeta, \label{eq:mapw}
\end{equation}
where ${\mathcal A}$ and $\mathcal K$ are complex constants, and the function $P(\zeta,q)$ is defined by
\begin{equation}
 P(\zeta,q)=(1-\zeta)\prod^{\infty}_{k=1}(1-q^{2k}\zeta)(1-q^{2k}\zeta^{-1}).
\end{equation}
Note that the four factors  of $P(\zeta/\alpha_i,q)$ under the square root in (\ref{eq:mapw}) ensure  that each point $\alpha_i$ is mapped to a right angle corner in the $z$-plane. Similarly, the points $\zeta=\beta_1$ and $\zeta=\beta_3$ are mapped to the slit endpoints in the $w$-plane, as implied by the fact that $dW/d\zeta=0$ at these two points. [To see this, note that $P(q^2,q)=0$]. 

The function  $P(\zeta,q)$ is related to the Jacobi elliptic functions but we shall not exhibit this relation here; see, e.g.,  \cite{crowdy1}.   For later use, we list the following properties of  $P(\zeta,q)$ that hold
for $|\zeta|=1$:
\begin{align}
\overline{{P({\zeta},q)}} &=-{\frac{1}{\zeta}}{P(\zeta,q)}, 
  \label{eq:aux01}
\end{align}
\begin{align}
\overline{P(q\zeta,q)} &= P(q\zeta,q), 
 \label{eq:aux02} 
\end{align}
\begin{align}
 \overline{P(q^2\zeta,q)} = -\zeta P(q^2\zeta,q),
 \label{eq:aux03}
\end{align}
where the overline indicates complex conjugate.

To completely specify the solution for $W(\zeta)$  we need to determine the parameters appearing in Eq.~(\ref{eq:mapw}).  Without loss of generality, we can set ${\cal A}=0$  and $\zeta_0=\alpha_4$, since this merely fixes the origin. Now, recalling that the mapping function  between a doubly connected domain and an annulus is uniquely defined  up to a factor of modulus one \cite{book}, we can choose the constant $\mathcal K$ to be purely imaginary, i.e., $\mathcal K= i K$,  where $K$ is a real constant to be determined later.

Next,  we note that there should be four free parameters corresponding to the four physical parameters of our solution, namely, the length $L$ of the unit cell,  the centroid of the bubble, and the bubble area. The radius $q$ of the inner circle is one such free parameter, which mainly governs the bubble area. For convenience, we choose the other three free parameters to be the points $\alpha_1$, $\alpha_2$, and $\alpha_3$ on the unit circle. 
The  remaining point $\alpha_4$  
is then determined by imposing the condition that the rectangle in the $w$-plane has the proper orientation. This can be done by requiring, for instance, that the upper edge of the unit cell in the $w$-plane is horizontal, a condition that can be written as
\begin{equation}
\text{Im}\left[\frac{dW}{d\theta}\right]=0 \quad\mbox{on}\quad \zeta=\exp(i \theta), \quad\mbox{for}\quad\theta_2 < \theta < \theta_3,
\label{eq:Im1}
\end{equation}
where $\theta_i=\arg(\alpha_i)$.
Using the fact that  ${dW}/{d\theta}=i\zeta({dW}/{d\zeta})$ on $|\zeta|=1$,  Eq.~(\ref{eq:Im1}) becomes
\begin{equation}
 \frac{dW}{d\zeta} =  - \frac{1}{\zeta^{2}} \,\overline{\left(\frac{dW}{d\zeta}\right)} \quad\mbox{on}\quad \zeta=\exp(i \theta), \quad\mbox{for}\quad\theta_2 < \theta < \theta_3.
\end{equation}
Using Eq.~(\ref{eq:mapw}) and the relations (\ref{eq:aux01}) and (\ref{eq:aux02})  in the preceding equation, one obtains after straightforward algebra the following relation between the $\alpha_{i}$'s:
\begin{equation}
 { \alpha_1\alpha_2 \alpha_3 \alpha_4 }=1,\label{cond.w.02b}
\end{equation}
which can be solved for $\alpha_4$ in terms of the other three (freely specified) parameters: $\alpha_{4}=\overline{ \alpha_1\alpha_2 \alpha_3 }$, where we have used that $1/\alpha_{i}=\bar\alpha_{i}$, since $|\alpha_{i}|=1$.

Similarly, from the condition that  the slit in the $w$-plane 
is horizontal, one obtains the following relation:
\begin{align}
\text{Im}\left[\frac{dW}{d\theta}\right] = 0 \quad\text{for} \quad \zeta=qe^{i\theta},
\end{align}
or alternatively
\begin{equation}
\frac{dW}{d\zeta} = - \frac{q^2}{\zeta^{2}} \,\overline{\left(\frac{dW}{d\zeta}\right)}  \quad\text{for} \quad {\zeta=qe^{i\theta}},
\label{eq:bcn1}
\end{equation}
where we have used that ${dW}/{d\theta}=i\zeta({dW}/{d\zeta})$ and $\bar \zeta = q^{2}/\zeta$ for $\zeta=qe^{i\theta}$. Using Eqs.~(\ref{eq:mapw}), (\ref{eq:aux02}), and (\ref{eq:aux03}) in Eq.~(\ref{eq:bcn1}) yields the relation
\begin{equation}
 \beta_{1}=\overline{\beta_{3}} .\label{eq:w01b}
\end{equation}
We must also ensure single-valuedness of the function $W(\zeta)$ by applying the condition
\begin{equation}
\oint_{|\zeta|=q}\,W(\zeta)\,d\zeta = 0 .
\label{eq:w03}
\end{equation}
Eqs.~(\ref{eq:w01b}) and (\ref{eq:w03}) thus determine the parameters $\beta_{1}$ and $\beta_{3}$.
Finally, the value of $K$ is determined by the following condition:
\begin{equation}
 \Delta W|^{\rm D}_{\rm A}\equiv W(\alpha_4)-W(\alpha_3)= i  a(U-V), \label{cond.k}
\end{equation}
as follows from Fig.~\ref{fig:2b}.

\subsection{The function $\tilde W(\zeta)$}

Since the flow domain in the $\tilde w$-plane is a polygonal region with a vertical slit in its interior [see Fig.~\ref{fig:2c}], 
the function $\tilde W(\zeta)$ can also be obtained from the  Schwarz-Christoffel formula:
\begin{equation}
 \tilde W(\zeta)=i \tilde K\int_{\alpha_4}^\zeta \frac{P(q^2\zeta/\beta_2,q)P(q^2\zeta/\beta_4,q)}
{\sqrt{\prod_{i=1}^4P(\zeta/\alpha_i,q)}}\, d\zeta, \label{eq:maptw}
\end{equation}
where, as before,  we have chosen the pre-multiplying constant as purely imaginary, meaning that $\tilde K$ is a real constant. 

The requirement that the slit in the $\tilde w$-plane be vertical (see Fig.~\ref{fig:2c}) yields the following relation:
\begin{align}
\text{Re}\left[\frac{d\tilde W}{d\theta}\right] = 0 \quad\text{for} \quad \zeta=qe^{i\theta},
\end{align}
or alternatively
\begin{equation}
\frac{d\tilde W}{d\zeta} =  \frac{q^2}{\zeta^{2}} \,\overline{\left(\frac{d\tilde W}{d\zeta}\right)}  \quad\text{for} \quad {\zeta=qe^{i\theta}}.
\end{equation}
Now carrying out similar steps as those that led to Eq.~(\ref{eq:w01b}) from (\ref{eq:bcn1}), one finds that
\begin{equation}
 \beta_{2}  =-\overline{ \beta_{4}},\label{eq:wt01}
\end{equation}
which together with the  single-valuedness condition,
\begin{equation}
 \oint_{|\zeta|=q} \tilde{W}(\zeta)\,d\zeta = 0 , \label{eq:wt02}
\end{equation}
determine $\beta_{2}$ and $\beta_{4}$.  Lastly, the constant $\tilde K$  is  fixed by the condition
\begin{equation}
 \Delta \tilde W|^{\rm A}_{\rm D}\equiv\tilde W(\alpha_3)-\tilde W(\alpha_4)=i aV. \label{cond.k.lab}
\end{equation}

\section{Examples}
Without loss of generality, we shall set $a=1$ and $V=1$ throughout this section.  We also recall that solutions for any $U\ne2$ can be obtained  by a proper rescaling of the solutions with $U=2$ \cite{prsa2011}. We shall thus restrict ourselves to the case $U=2$.  

It is easy to verify that  the general solution presented above reproduces, as a particular case,  the solution obtained by Burgess and Tanveer \cite{BT} for an infinite stream of symmetrical bubbles. To see this, it suffices to choose the four points $\alpha_i$ to be symmetrically located with respect to the vertical and horizontal axes, i.e., $\alpha_{1}=-\bar \alpha_2=- \alpha_3=\bar \alpha_4$. One then finds  that $\beta_1=-\beta_3=i$ and  $\beta_2=-\beta_4=-1$, implying that the bubble is symmetric with respect both to the channel centerline and to the fore-and-aft direction, as in the Burgess-Tanveer solution \cite{BT}.  

 \begin{figure}[t]
 \centering
 \includegraphics[width=5cm]{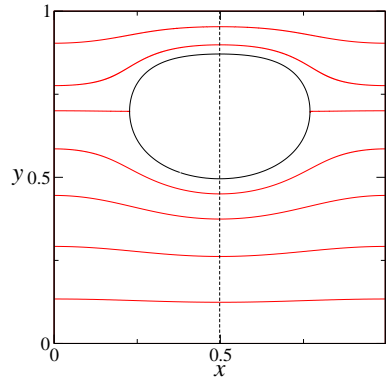}
 \caption{Period cell for a stream of bubbles  with fore-and-aft symmetry. Here the parameters are $\theta_1=0.3$, $\theta_2=1.64$, and $q=0.5$, and the resulting period is $L=0.996$ }
 \label{fig:3}
 \end{figure}

\begin{figure}[b]
\vspace{0.6cm}
\centering
\includegraphics[width=5cm]{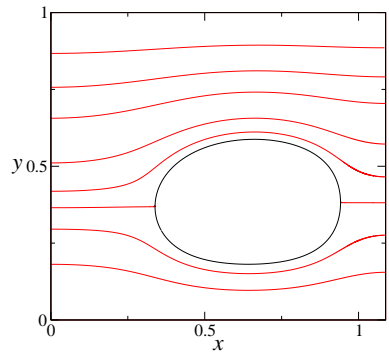}
\caption{Asymmetric solution for $\theta_1$=0.87, $\theta_2$=3.0, $\theta_3=3.9$, and $q$=0.5. The resulting half-period is $L=1.09$.}
\label{fig:4}
\end{figure}

If one requires instead only that   $\alpha_4=\bar \alpha_1$ and $\alpha_3=\bar{\alpha}_2$, with  no assumed relation between $\alpha_1$ and $\alpha_2$, one obtains a stream of bubbles with fore-and-aft symmetry but no centerline symmetry.  An example of this case is shown in Fig.~\ref{fig:3} for $\theta_1=0.3$, $\theta_2=1.64$, and $q=0.5$. (In this figure some streamlines of the flow are also shown for illustration purposes.) We note, in particular, that in the limit that the cell period goes to infinity, i.e., $L\to\infty$, the family of  solutions shown in Fig.~\ref{fig:3}  reproduces the exact solutions for a single asymmetric bubble in a Hele-Shaw cell  \cite{tanveer87,CD1988}. The limit $L\to\infty$ is accomplished in our formulation by setting $\alpha_{1}=\alpha_{2}$ and $\alpha_{3}=\alpha_{4}$; see Fig.~\ref{fig:2}.

Similarly, a stream of bubbles with centerline symmetry (but not with fore-and-aft symmetry) is obtained by imposing the conditions: $\alpha_1=-\bar \alpha_2$ and $\alpha_3=-\bar \alpha_4$. It is worth pointing out that solutions with either centerline or fore-and-aft symmetry can also be  obtained  by reducing the problem to a simply connected domain and applying the standard Schwarz-Christoffel formula  \cite{glv1994,prsa2011}. The formalism presented here, however, is more general in  that it naturally accounts for asymmetric bubbles. Indeed, if one chooses the parameters $\alpha_i$ arbitrarily, i.e., with no symmetry relations between them, then the resulting bubble has no symmetry whatsoever. One example of such solutions is shown in Fig.~\ref{fig:4} for the parameters  $\theta_1$=0.87, $\theta_2$=3.0, $\theta_3=3.9$, and $q$=0.5. 

\section{Conclusions}

We have presented an exact solution for a steady stream  of  asymmetric bubbles co-travelling in a Hele-Shaw channel. From the periodicity along the streamwise direction, the original flow domain was reduced to a rectangular unit cell containing only one bubble. The complex potentials $w(z)$ and $ \tilde w(z)$ in the co-moving frame and in the lab frame, respectively, were computed in terms of conformal mappings from an annulus in an auxiliary complex $\zeta$-plane onto the respective flow domains in the $w$- and $\tilde w$-planes.  Both these domains turn out to be given by degenerate polygonal regions, so that the desired mappings could be constructed explicitly by making use of the Schwarz-Christoffel formula for doubly connected domains. 
As an important extension of the present work it would be interesting to consider  periodic solutions with  several (asymmetric) bubbles per unit cell, in which case one needs to employ the generalized Schwarz-Christoffel formula for polygonal domains of arbitrary connectivity. This more difficult problem is currently under investigation.

\end{document}